# Unconventional superconductivity in the BiS$_2$-based layered superconductor NdO$_{0.71}$F$_{0.29}$BiS$_2$

Yuichi Ota[1], Kozo Okazaki[1,*], Haruyoshi Q. Yamamoto[1], Takashi Yamamoto[1], Shuntaro Watanabe[2], Chuangtian Chen[3], Masanori Nagao[4,5], Satoshi Watauchi[4], Isao Tanaka[4], Yoshihiko Takano[5,6], and Shik Shin[1,*]

[1]*Institute for Solid State Physics (ISSP), University of Tokyo, Kashiwa, Chiba 277-8581, Japan*
[2]*Research Institute for Science and Technology, Tokyo University of Science, Chiba 278-8510, Japan*
[3]*Beijing Center for Crystal R&D, Chinese Academy of Science (CAS), Zhongguancun, Beijing 100190, China*
[4]*Center for Crystal Science and Technology, University of Yamanashi, Kofu 400-8511, Japan*
[5]*National Institute for Materials Science, Tsukuba, Ibaraki 305-0047, Japan*
[6]*University of Tsukuba, Graduate School of pure and Applied Sciences, Tsukuba, Ibaraki 305-8577, Japan*
[*]*To whom correspondence should be addressed. E-mail: okazaki@issp.u-tokyo.ac.jp; shin@issp.u-tokyo.ac.jp*

(Dated: March 23, 2017)

We investigate the superconducting-gap anisotropy in one of the recently discovered BiS$_2$-based superconductors, NdO$_{0.71}$F$_{0.29}$BiS$_2$ ($T_c \sim 5$ K), using laser-based angle-resolved photoemission spectroscopy. Whereas the previously discovered high-$T_c$ superconductors such as copper oxides and iron-based superconductors, which are believed to have unconventional superconducting mechanisms, have 3$d$ electrons in their conduction bands, the conduction band of BiS$_2$-based superconductors mainly consists of Bi 6$p$ electrons, and hence the conventional superconducting mechanism might be expected. Contrary to this expectation, we observe a strongly anisotropic superconducting gap. This result strongly suggests that the pairing mechanism for NdO$_{0.71}$F$_{0.29}$BiS$_2$ is unconventional one and we attribute the observed anisotropy to competitive or cooperative multiple paring interactions.

Since the discovery of BiS$_2$-based superconductor LaO$_{1-x}$F$_x$BiS$_2$ by Mizuguchi *et al.* [1], this class of superconductors has attracted much attention and several new members were found soon after that [2–6]. They have several features similar to those of the previously discovered high-$T_c$ superconductors of copper oxides (cuprates) [7] and iron-based superconductors [8]; that is, they have a layered crystal structure, and in LnO$_{1-x}$F$_x$BiS$_2$ (Ln = lanthanoid), superconductivity emerges at various Ln contents [9] with electron doping caused by substitution of F for O like the iron-pnictide LnFeAsO$_{1-x}$F$_x$ (1111 system). Hence, the mechanism of superconductivity in BiS$_2$-based superconductors may be expected to resemble that of the iron-pnictides. However, in contrast to the other high-$T_c$ superconductors, the conduction bands of BiS$_2$-based superconductors consist mainly of Bi 6$p$ electrons [10], which is expected to have relatively weak electronic correlations, rather than 3$d$ electrons, which have strong correlations that had been considered as indispensable for high-$T_c$ superconductivity. Hence, in BiS$_2$-based superconductors, conventional electron-phonon coupling might be considered to be dominant for their superconductivity.

However, recent neutron scattering experiments have indicated that the electron-phonon coupling is much weaker than expected from the above scenario [11], and suggested the importance of charge fluctuations to superconductivity in BiS$_2$-based superconductors [12]. In addition, a large $2\Delta/k_B T_c$ suggests that the pairing mechanism is unconventional [13, 14] and theoretical studies have suggested unconventional superconducting (SC) pairing mechanisms on the basis of calculations that assume superconductivity driven purely by electron-electron coupling [10, 15–18]. These studies have indicated that the Fermi surface (FS) topology and SC gap anisotropy depend strongly on carrier doping; hence, it is crucial for understanding the superconducting mechanism to directly observe the band structure and SC gap anisotropy in these compounds.

Angle-resolved photoemission spectroscopy (ARPES) is a powerful tool for direct observation of the electronic structure and SC gap [19], and it has already been used to reveal the basic electronic structure of BiS$_2$-based superconductors [20, 21]. For NdO$_{1-x}$F$_x$BiS$_2$, which has a maximum $T_c$ of 5.6 K [4], FSs smaller than those expected from the nominal doping have been reported [20, 21]. Direct observation of the SC gap in low-$T_c$ materials with $T_c$ as low as $\sim 3$ K has recently become possible by using a laser-ARPES apparatus that achieves the highest energy resolution of 70 $\mu$eV and the lowest temperature of 1.5 K. Using this apparatus, we measured the band structure and SC gap anisotropy of a 29 % F-doped sample (NdO$_{0.71}$F$_{0.29}$BiS$_2$, $T_c \sim 5$ K) and found highly anisotropic SC gaps. By carefully adjusting the focal point of the laser and reducing the spot size to as small as $\sim 100$ $\mu$m at the cleaved sample surface, we can probe a region of few defects and detect a highly anisotropic and possibly nodal SC gap structure. We conclude that our result strongly suggests that NdO$_{0.71}$F$_{0.29}$BiS$_2$ is an unconventional superconductor that has competitive or cooperative multiple pairing interactions.

High quality single crystals of NdO$_{0.71}$F$_{0.29}$BiS$_2$ were grown by a CsCl/KCl flux as described in Ref. [22]. The amount of F substitution for O was determined from electron probe microanalysis (EPMA) measurements. Clean surfaces were obtained by cleaving the sample *in situ* under ultrahigh vacuum (better than $5 \times 10^{-11}$ Torr). ARPES measurements were performed using the He I$\alpha$ ($h\nu$ = 21.218 eV) resonance line with a VG-Scienta R4000 electron analyzer and using a VUV laser ($h\nu$ = 6.994 eV) with a VG-Scienta HR8000 electron analyzer [23]. To minimize the space charge effect, the repetition rate of the VUV laser was raised to 960 MHz. The average power of the VUV laser is in the order of 10 $\mu$W and does not affect cleanliness of the sample surfaces. The total energy resolution was set to 15 meV for the measurements using the He discharge lamp and to $\sim$ 1.2 and 4 meV for the SC gap measurements and FS and $E$-$k$ map measurements, respectively, using the VUV laser. More details for the laser ARPES apparatus have been described in literature [23–25]. The SC transition temperature was confirmed by magnetization measurements before the ARPES measurements of the same samples.

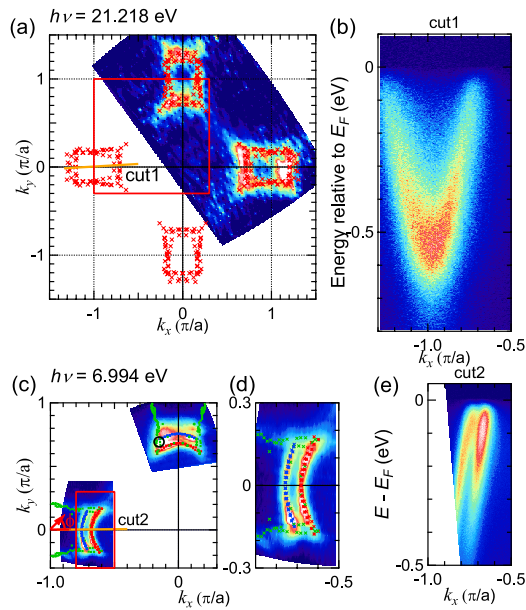

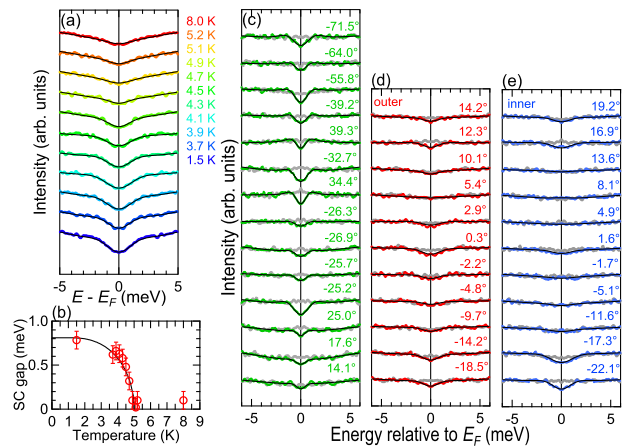

FIG. 1. FS and $E$-$k$ maps of NdO$_{0.71}$F$_{0.29}$BiS$_2$. FS maps measured using (a) He I$\alpha$ at 10 K and (c) a VUV laser at 8 K. The integration energy window of each FS map is ±5 meV from $E_F$. Red, blue, and green symbols indicate the positions of $k_F$. The solid square in panel (a) corresponds to the region displayed in panel (c). Definition of the FS angle $\phi$ is shown in panel (c), and the open circle in panel (c) indicates the $k_F$ for the temperature-dependent EDC shown in Fig. 2(a). (b) $E$-$k$ map along cut 1 in panel (a) measured using He I$\alpha$ at 10 K. (d) Enlarged plot for the region indicated by the solid square displayed in panel (c). (e) $E$-$k$ map along cut 2 in panel (c) measured using the VUV laser at 8 K.

FIG. 2. Temperature- and FS-angle-dependent symmetrized EDCs at $k_F$ measured using $c$-polarized light. (a) Temperature-dependent EDC symmetrized with respect to $E_F$ at $k_F$ indicated by the open circle in Fig. 1(c). (b) Temperature dependence of the SC gap size deduced from the fitting. The solid line indicates the SC gap size from the BCS theory for $\Delta(0) = 810$ $\mu$eV and $T_c = 5$ K. (c)-(e) Symmetrized EDCs at $k_F$ (c) in the degenerate region, and clearly separated region of (d) outer FS and (e) inner FS. The EDCs were measured at 1.5 and 8 K ($T_c \sim 5$ K) for various FS angles $\phi$, as shown in each panel. Gray symbols indicate the EDCs above $T_c$ and solid lines indicate the fitting functions using a BCS spectral function. EDCs before symmetrization are shown in Fig. S2 [26].

First, we determined the FS topology and the positions of the Fermi momentum ($k_F$) for all the FSs. Figures 1(a) and 1(c) show the FS maps measured using the He I$\alpha$ line at 10 K and using a vacuum ultraviolet (VUV) laser with circular ($c$) polarization at 8 K, respectively, and Fig. 1(d) is the enlarged plot of the region indicated by the solid square in Fig. 1(c). The $k_F$ positions of the FSs were determined by fitting the momentum distribution curves (MDCs) to Lorentzians and are indicated by the symbols on the FS maps in Figs. 1(a), 1(c), and 1(d). The FS maps were created from the MDCs integrated within ± 5meV of $E_F$. Figures 1(b) and 1(e) show the $E$-$k$ maps measured using the He I$\alpha$ line and the VUV laser along cuts 1 and 2 shown in Figs. 1(a) and 1(c), respectively. The $E$-$k$ map in Fig. 1(b) clearly shows that the FSs are composed of an electron-like band, whereas that in Fig. 1(e) shows that two electron bands are separated along this momentum cut [27]. Two FSs are recognized in this region, as indicated by the red and blue symbols in Fig. 1(c) and 1(d), and we call these FS sheets the outer and inner FSs, respectively. However, in the region indicated by the green symbols, two FS sheets are almost degenerate and cannot be separated. We call this region the degenerate region. The outer and inner FSs enclose 6.3% and 5.0% of the Brillouin zone area, respectively, and this corresponds to electron doping of 22.6% per Bi site. This is almost consistent with the electron doping expected from the amount of F substitution determined from EPMA, and in contrast to the observation that the FS volume was considerably smaller than that expected from the nominal amount of F substitution for $NdO_{0.7}F_{0.3}BiS_2$ by Zeng et al. [21] This may be attributable to the better quality of our samples or bulk sensitivity of our measurements. Additionally, the shape of the FSs differs somewhat from that previously observed, and it is winding in the degenerate region and curved in the separation region. This should be also originated from the fact that the amount of F substitution for our samples is larger than that for the samples measured by Zeng et al.

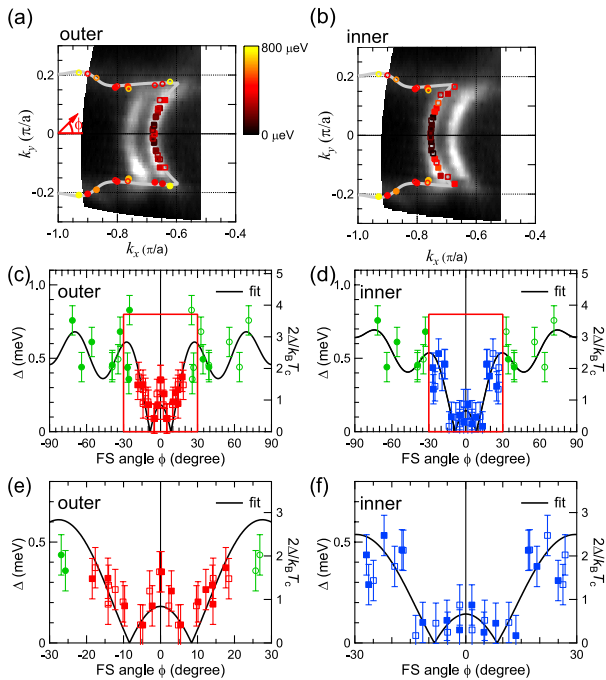

FIG. 3. SC gap anisotropy of $NdO_{0.71}F_{0.29}BiS_2$. (a), (b) The $k_F$ positions at which the SC gap was measured plotted on the FS map image for the outer and inner FSs, respectively. Color scale corresponds to each SC gap size derived from the fitting. Solid symbols are derived from the EDCs shown in Fig. 2; open symbols are plotted after symmetrization, taking into account the tetragonal crystal symmetry. (c), (d) SC gap sizes derived from the fitting plotted as a function of FS angle for the outer and inner FSs, respectively. Error bar denotes the standard deviation of the $E_F$ position (see Supplemental Material and Fig. S6 [26]). Solid lines are fitting functions using a model gap function of $\Delta(\phi) = |\Delta_0(1 + \Delta_2 \cos(2\phi) + \Delta_4 \cos(4\phi) + \Delta_6 \cos(6\phi) + \Delta_8 \cos(8\phi))|$. (e), (f) Enlarged plots for the regions indicated by solid squares in panels (c) and (d), respectively.

Next, to reveal the SC gap structure of $NdO_{0.71}F_{0.29}BiS_2$, we measured the SC gaps on the two electron FS sheets using the VUV laser. The energy distribution curves (EDCs) symmetrized with respect to $E_F$ are shown in Fig. 2 [28]. Figure 2(a) shows the temperature-dependent symmetrized EDCs at $k_F$ indicated by the open circle in Fig. 1(c). A valley structure at $E_F$ indicates the existence of the SC gap, and it vanishes at a temperature close to $T_c$ within a narrow temperature range (4.9-5.1 K) reflecting a clear transition between the normal and superconducting phases. We quantified the SC gap size by fitting the EDCs to the Bardeen-Cooper-Schrieffer (BCS) spectral function as shown in Fig. S2, and the deduced temperature-dependent SC gap size $\Delta(T)$ is shown in Fig. 2(b). More detailed analyses to ensure the validity of our evaluation of the SC-gap size are described in Supplemental Material [26]. The solid line indicates the temperature dependence of the SC gap size based on the BCS theory for $\Delta(0) = 810$ $\mu$eV and $T_c = 5$ K, corresponding to the reduced gap size $2\Delta(0)/k_B T_c = 3.76$. One can confirm that the temperature dependence of the deduced SC gap size is in accordance well with that of the BCS theory. Figures 2(c)-(e) show the symmetrized EDCs at $k_F$ in the degenerate region and the clearly separated regions of the outer and inner FSs, respectively, measured at 1.5 K (below $T_c$) and 8 K (above $T_c$). Each symmetrized EDC at $k_F$ is identified with an FS angle $\phi$ for each FS [Fig. 1(c)]. The valley structures at $E_F$ observed in the symmetrized EDCs of both FSs [Figs. 2(c)-(e)] indicate the opening of the SC gap, and the flat structure indicates that the SC gap is quite small. For both FSs, the line shape of the symmetrized EDCs at $k_F$ depends strongly on the FS angle, indicating that the SC gap is highly anisotropic.

The SC gap anisotropy deduced from the fitting is shown in Fig. 3. The right axes of Figs. 3(c) and 3(d) correspond to $2\Delta/k_B T_c$, and its maximum is comparable to the BCS value of 3.53. This suggests that our results are reliable but is in contrast to previous reports [13, 14]. The open symbols are plotted after symmetrization, taking into account the tetragonal crystal symmetry. Around $\phi = 5°$ for the outer FS and $\phi = 8°$ for the inner FS, $\Delta$ shows a cusp-like minimum, whereas it is clearly finite at $|\phi| > 15°$. This is also clear from the symmetrized EDCs (see also Fig. S5). They show flat structures around $\phi = 5°$ for the outer FS and $\phi = 8°$ for the inner FS, whereas they clearly show valley structures at $E_F$ at $|\phi| > 15°$. This indicates the existence of SC gap nodes. Considering the tetragonal crystal symmetry, the FSs around the X point have twofold rotational symmetry, and their SC gap anisotropy should also have twofold symmetry with respect to the X point. Hence, we can use the following function to fit the SC gap anisotropy.

$$\Delta(\phi) = |\Delta_0(1 + \Delta_2 \cos(2\phi) + \Delta_4 \cos(4\phi) \\ + \Delta_6 \cos(6\phi) + \Delta_8 \cos(8\phi))|, \quad (1)$$

for each FS. Although fairly high-order terms are needed for the fitting because the FSs are rectangular rather than ellipsoidal, this fitting function successfully reproduces the SC gap anisotropy of both FSs as shown in Fig. 3.

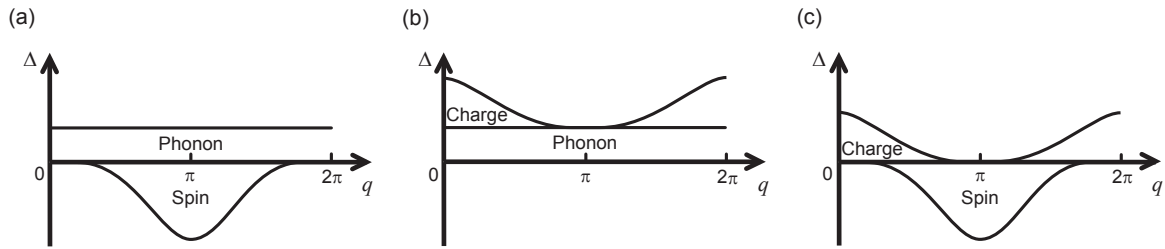

FIG. 4. Schematic illustration of multiple pairing interactions. Positive and negative SC gaps $\Delta$ correspond to those originated from attractive and repulsive interactions, respectively. (a) Competition of $q$-independent conventional electron-phonon coupling and $q$-dependent repulsive spin fluctuations. (b) Cooperation of $q$-independent electron-phonon coupling and $q$-dependent attractive charge fluctuations. (c) Competition of $q$-dependent attractive charge fluctuations and $q$-dependent repulsive spin fluctuations. The attractive charge fluctuations are assumed to be peaked at $q = (0,0)$, whereas the repulsive spin fluctuations are assumed to be peaked at $q = (\pi,\pi)$.

Here, we focus on the SC gap anisotropy and origin of the pairing interactions. As shown in Fig. 3, the SC gaps of the outer and inner FSs show strong anisotropies, and we determined that they have node-like minima. According to random phase approximation calculations, a similar FS topology can produce a $d$- or $g$-wave SC gap symmetry and hence a nodal SC gap anisotropy, and its anisotropy is similar to our results that the nodes are located at the short edge of the rectangular FSs [10, 18, 29]. On the other hand, magnetic penetration depth measurements and thermal transport measurements have indicated nodeless superconductivity in NdO$_{1-x}$F$_x$BiS$_2$ ($x$ = 0.3 and 0.5) [30, 31]. These results may seem to be inconsistent with our results. However, if the SC gap symmetry of NdO$_{1-x}$F$_x$BiS$_2$ is $s$-wave and the SC gap nodes are accidental ones, the nodes can be lifted by disorder effects because they are not symmetry-protected in nodal $s$-wave superconductivity. Such a disorder-induced topological change of the SC gap structure from a nodal $s$-wave to a nodeless $s$-wave has been reported for iron-pnictide superconductors [32–34]. For NdO$_{1-x}$F$_x$BiS$_2$, the existence of bismuth and sulfur defects has been reported from several experiments [20, 21, 35], and this could also change the SC gap structure from the nodal $s$-wave to the nodeless one in bulk measurements. On the other hand, ARPES is a surface-sensitive technique, although laser ARPES using a VUV laser is relatively bulk-sensitive, and the VUV laser can be focused to a spot size of $\sim$ 100 $\mu$m at the cleaved sample surface by carefully adjusting the focal point of the laser. With this surface sensitivity and small spot size, laser ARPES can probe a region of few defects and detect the highly anisotropic and possibly nodal SC gap structure [36]. In this work, without careful adjustment of the focal point of the laser, we observed an anisotropic but nodeless SC gap, and the sufficiently reproduced data could not be obtained. Because the SC gap anisotropy measured by ARPES is generally suppressed due to the existence of impurities and/or disorders [37], the observed anisotropy should reflect the region free from defects and/or disorders. In addition, if the observed anisotropy is originated from defects and/or disorders, such anisotropy cannot be reproduced because defects and/or disorder should be different for piece by piece or sample positions, and hence, the obtained reproducibility assures that the observed anisotropy is intrinsic.

As mentioned above, our results revealed a strong SC gap anisotropy and suggested the existence of accidental nodes in nodal $s$-wave symmetry. This should indicate competition or cooperation among multiple pairing interactions. First, a strong SC gap anisotropy suggests contributions from unconventional SC pairing interactions that have a strong $q$-dependence. On the other hand, a finite $q$-independent component $\Delta_0$ can be expected to originate from the conventional pairing interaction of phonons. Thus, we assumed competition and cooperation among three types of SC-pairing interactions as the typical possible origins of superconductivity in NdO$_{0.71}$F$_{0.29}$BiS$_2$: (i) $q$-independent electron-phonon coupling, which produces conventional superconductivity, (ii) $q$-dependent repulsive spin fluctuations, and (iii) $q$-dependent attractive charge fluctuations. As schematically shown in Fig. 4, we considered competition and cooperation among these three interactions as the origin of the SC gap anisotropy in NdO$_{0.71}$F$_{0.29}$BiS$_2$. Here, positive and negative SC gaps $\Delta$ correspond to those originated from attractive and repulsive interactions, respectively. In Fig. 4(a), $q$-independent electron-phonon coupling and $q$-dependent spin fluctuations are assumed. Whereas electron-phonon coupling is an attractive interaction, spin fluctuations are originated from a repulsive Coulomb interaction; thus, these interactions should be competitive. In this case, depending on the ratio of these contributions, the SC gap can be nodal. On the other hand, in Fig. 4(b), $q$-independent electron-phonon coupling and $q$-dependent attractive charge fluctuations are assumed. In this case, the two interactions are cooperative; thus, the SC gap can be strongly anisotropic but nodeless. In Fig.4(c), $q$-dependent repulsive spin fluctuations and attractive charge fluctuations are assumed. Both interactions are anisotropic and the SC gap likely becomes nodal. We cannot conclude which case is the most appropriate for $NdO_{0.71}F_{0.29}BiS_2$, but in any case, it is strongly suggested that the mechanism of superconductivity in $NdO_{0.71}F_{0.29}BiS_2$ is unconventional.

In summary, we investigated the band structure and SC gap anisotropy of $NdO_{0.71}F_{0.29}BiS_2$ by ARPES measurements using a He discharge lamp and a VUV laser. We observed two electron FSs and found that their SC gaps are strongly anisotropic and have node-like minima. Whereas the observed SC gap anisotropy is similar to that of the theoretical results suggesting a possibility of $d$- or $g$-wave for the SC gap symmetry of $NdO_{0.71}F_{0.29}BiS_2$, $s$-wave with accidental nodes is strongly suggested from the comparison to the other experimental results. We considered competition and cooperation of multiple pairing interactions and concluded that $NdO_{0.71}F_{0.29}BiS_2$ is an unconventional superconductor.

We thank R. Arita, K. Kuroki, H. Usui, K. Suzuki, and T. Hotta for valuable discussions and comments. This work was supported by the Photon and Quantum Basic Research Coordinated Development Program of MEXT and partially supported by JSPS KAKENHI Grant Numbers JP25220707 and JP26610095.

been reported for cuprates [38] and iron-based superconductors [39], and thus disorder and/or defects also could be in this scale.

# Supplemental Material for "unconventional superconductivity in the BiS$_2$-based layered superconductor NdO$_{0.71}$F$_{0.29}$BiS$_2$"

Yuichi Ota[1], Kozo Okazaki[1,*], Haruyoshi Q. Yamamoto[1], Takashi Yamamoto[1], Shuntaro Watanabe[2], Chuangtian Chen[3], Masanori Nagao[4,5], Satoshi Watauchi[4], Isao Tanaka[4], Yoshihiko Takano[5,6], and Shik Shin[1,*]

[1]*Institute for Solid State Physics (ISSP),*
*University of Tokyo, Kashiwa, Chiba 277-8581, Japan*
[2]*Research Institute for Science and Technology,*
*Tokyo University of Science, Chiba 278-8510, Japan*
[3]*Beijing Center for Crystal R&D, Chinese Academy of Science (CAS),*
*Zhongguancun, Beijing 100190, China*
[4]*Center for Crystal Science and Technology,*
*University of Yamanashi, Kofu 400-8511, Japan*
[5]*National Institute for Materials Science, Tsukuba, Ibaraki 305-0047, Japan*
[6]*University of Tsukuba, Graduate School of pure and*
*Applied Sciences, Tsukuba, Ibaraki 305-8577, Japan*
*To whom correspondence should be addressed. E-mail:
okazaki@issp.u-tokyo.ac.jp; shin@issp.u-tokyo.ac.jp*

(Dated: March 23, 2017)

# MOMENTUM DISTRIBUTION CURVE AND EVALUATION OF MEAN FREE PATH

Peak width of a momentum distribution curve (MDC) reflects the mean free path $l$ of quasiparticles and it can be shortened by impurity scattering. Hence from the analysis of MDCs, cleanliness of the measured surface can be evaluated. Figure S1(b) shows a MDC and an $E$-$k$ map along the cut1 indicated in Fig. S1(a). The integration energy window of the MDC is $\pm$ 5 meV from $E_F$. From the fitting to Lorentzian curves, we have estimated the full width at half maximum (FWHM) of the MDC peaks ($\Delta k$) to be $\sim 0.08$ $\mathring{A}^{-1}$. Because the inverse half width at half maximum of a MDC corresponds the mean free path [1], the mean free path $l$ is estimated to be 25 $\mathring{A}^{-1}$ and this is several times longer than the in-plane lattice constant $a \sim 3.96$ $\mathring{A}^{-1}$. This indicates that the quasiparticles observed in this spectrum are enough coherent and not so much disturbed by impurities and/or inhomogeneity. Compared to iron-pnictides, of which SC gap structure has been actively discussed from various experiments including photoemission spectroscopy, the MDC peak width of $BaFe_2(As_{1-x}P_x)_2$, which shows a fairly high residual resistivity ratio (RRR) of several hundreds and known as one of the cleanest systems among the iron-pnictides [2], is comparable or broader than that of $NdO_{0.71}F_{0.29}BiS_2$ shown in Fig. S1(b) [3–5]. Also for $Ba_{1-x}K_xFe_2As_2$, of which end material $KFe_2As_2$ also shows several hundreds of RRR, a detailed doping dependence of superconducting-gap structure, which is consistent with bulk measurements of magnetic penetration depth or thermal conductivity, has been reported [6, 7] and their MDC widths are actually broader than that of $NdO_{0.71}F_{0.29}BiS_2$. From these facts, we have concluded that spectra of $NdO_{0.71}F_{0.29}BiS_2$ shown here have enough quality to determine the detailed and intrinsic momentum dependence of the superconducting gap.

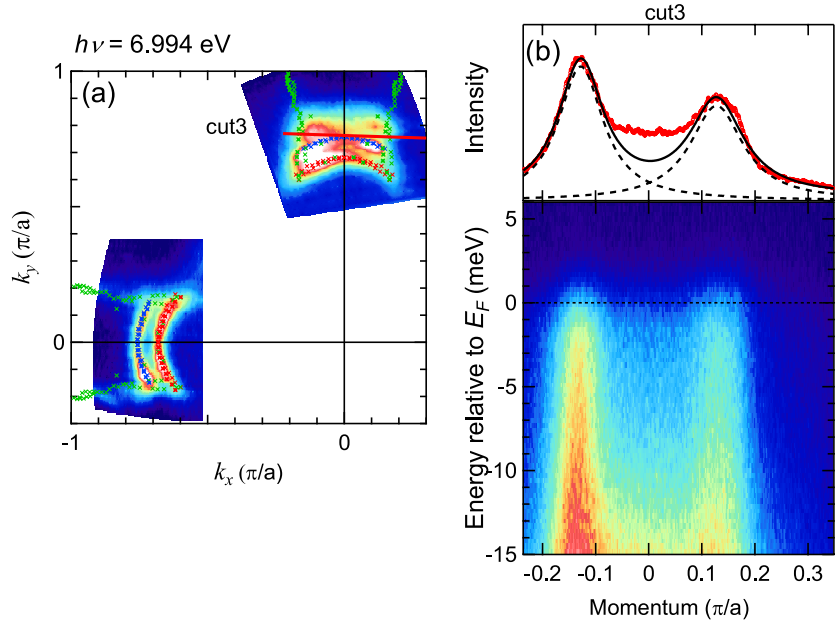

FIG. S1. MDC analysis for NdO$_{0.71}$F$_{0.29}$BiS$_2$. (a) FS maps taken with a VUV laser at 8 K. (b) MDC and $E$-$k$ maps along the cut3 in the panel (a).

# TEMPERATURE- AND FS-ANGLE-DEPENDENT EDCS BEFORE SYMMETRIZATION AND FITTING TO THE BCS SPECTRAL FUNCTION.

Figure S2 shows the EDCs before symmetrization of which symmetrized ones are shown in Fig. 2. In order to quantify the SC gap size, we fitted the EDCs to the Bardeen-Cooper-Schrieffer (BCS) spectral function of the form [6, 8],

$$A_{\rm BCS}(k,\omega) = \frac{1}{\pi}\left\{\frac{|u_k|^2\Gamma}{(\omega-E_k)^2+\Gamma} + \frac{|v_k|^2\Gamma}{(\omega+E_k)^2+\Gamma}\right\},\ E_k = \sqrt{\epsilon_k^2 + |\Delta(k)|^2},$$

where $|u_k|^2$ and $|v_k|^2$ are coherence factors, $\Gamma$ is a broadening factor due to the finite quasiparticle lifetime, $E_k$ and $\epsilon_k$ are Bogoliubov-quasiparticle and bare-electron energies, respectively, and $\Delta(k)$ is a SC gap magnitude. All the EDCs shown in Fig. S2 could be fitted to the calculated spectra, as indicated by the solid lines.

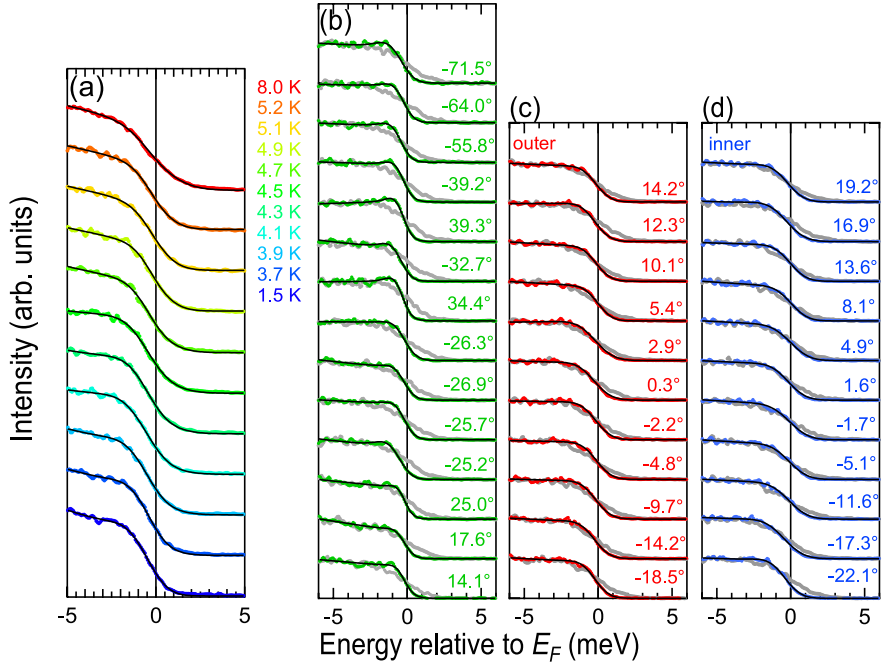

FIG. S2. Temperature- and FS-angle-dependent EDCs before symmetrization. (a) Temperature-dependent EDCs at $k_F$ indicated by the open circle in Fig. 1(c). (b)-(d) EDCs at $k_F$ (b) in the degenerate region, and clearly separated region of (c) outer FS and (d) inner FS, respectively.

**SHIFT OF THE LEADING EDGE MIDPOINT**

We evaluated the temperature dependence of the SC gap also from the shift of the leading edge midpoint of the EDCs at $k_F$. Figure S3(a) shows the EDCs shown in Fig. S2(a) without offset. The region bounded by the square in Fig. S3(a) is enlarged in Fig. S3(b) to show the shift of the leading edge midpoint (LEM). A systematic shift to the lower energy can be recognized below 4.9 and it is consistent with the bulk $T_c$. In Fig. S3(c), we have plotted the temperature dependence of the LEM shift (left axis) and the temperature dependence of the SC gap based on the BCS theory for $\Delta(0) = 810$ $\mu$eV and $T_c = 5$ K (right axis). It can be confirmed that the LEM shift is proportional to the SC gap and the SC gap is fully open at the measured temperature of 1.5 K.

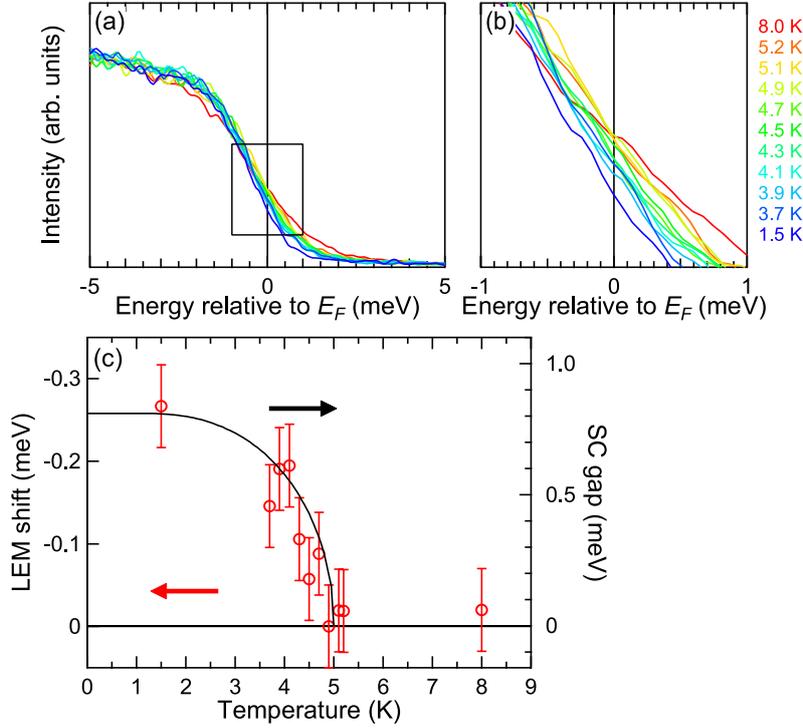

FIG. S3. Shift of the leading edge midpoint. (a) Temperature dependence of the EDCs at $k_F$, which are the same as those shown in Fig. S2(a) but plotted without offset. (b) Enlarged plot of the region indicated by a square in panel (a) to show the LEM shift. (c) Temperature dependence of the LEM shift (left axis) and the temperature dependence of the SC gap size from the BCS theory (right axis), the same as shown in Fig. 2(b).

# TEMPERATURE VARIATION OF EDC ABOVE AND BELOW $T_c$ AND SC COHERENCE PEAK

In order to extract the intrinsic temperature variation of the EDC, we show the temperature dependence of the symmetrized EDCs divided by that above $T_c$ (8 K) in Fig. S4(a). As indicated by the arrows, the superconducting coherence peaks can be recognized and it supports that the observed temperature dependence of the EDCs reflect bulk superconductivity. We have averaged the EDCs above $T_c$ and the EDCs below $T_c$, respectively, and divided the averaged EDC below $T_c$ by the averaged EDC above $T_c$. The result is shown in Fig. S4(b) with the enlarged vertical axis. The superconducting coherence peaks can be recognized more clearly.

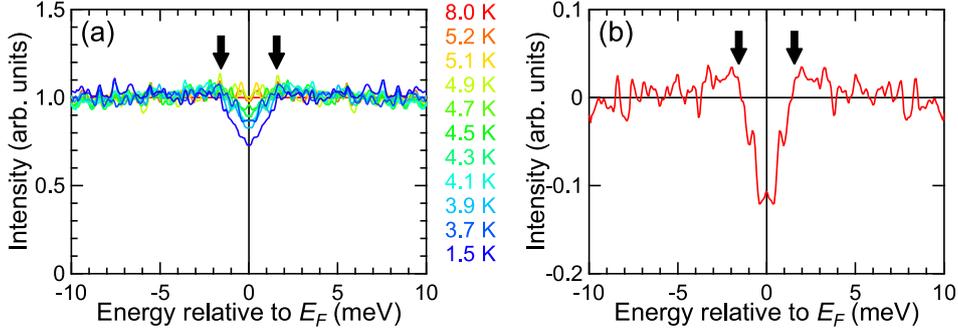

FIG. S4. (a) Symmetrized EDCs divided by EDC above $T_c$ (8K). (b) the averaded EDC below $T_c$ divided by the averaged EDC above $T_c$. The $k_F$ position for these EDCs is indicated by the open circle in Fig. S1(a). The arrows indicate a coherence peak.

# SUPPRESSION OF THE INTENSITY AT $E_F$ BELOW $T_c$

The reduction of the intensity at $E_F$ of the symmetrized EDCs below $T_c$ is shown in Fig. S5 (left axis) with the SC gap anisotropy (right axis). We can see that suppression of the intensity at $E_F$ below $T_c$ strongly correlate with the gap opening, and the existence of the SC gap nodes are strongly suggested.

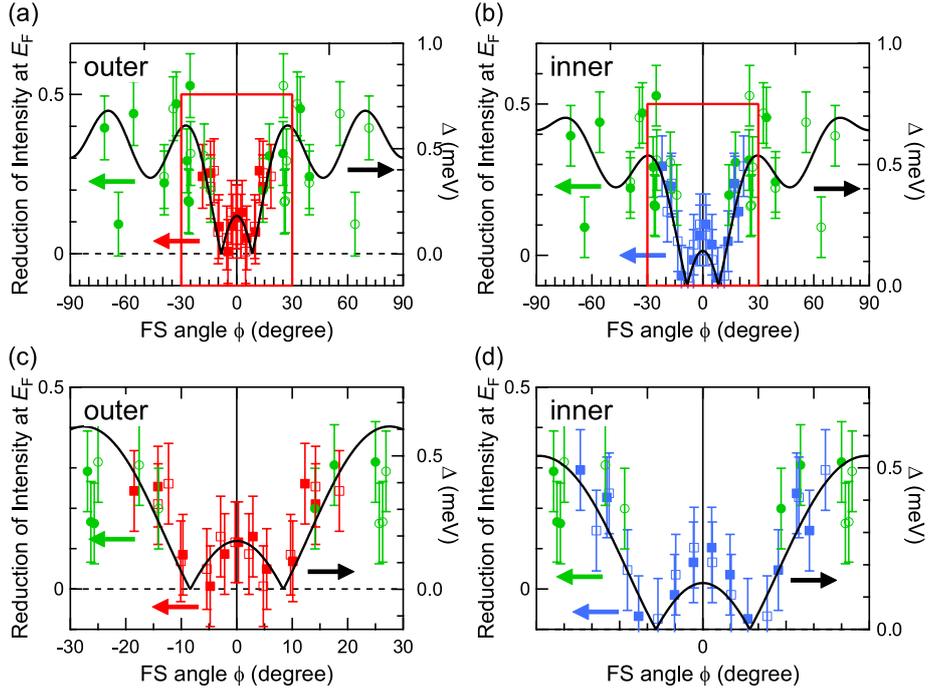

FIG. S5. Suppression of the intensity at $E_F$ of the symmetrized EDCs below $T_c$. (a), (b) Reduction of the intensity at $E_F$ below $T_c$ relative to that above $T_c$ $[I(8\text{K}) - I(1.5\text{K})]/I(8\text{K})$, where $I(T\text{K})$ is the intensity at $E_F$ of the symmetrized EDC at $T$ K, plotted as a function of FS angle for the outer and inner FSs, respectively (left axis). The black lines are the fitting results for the SC gap anisotropy shown in Figs. 3(c)-(f) (right axis). (c), (d) Enlarged plots for the regions indicated by red squares in panels (a) and (b), respectively.

## STABILITY OF $E_F$ POSITION

Because we have discussed the size of superconducting gap smaller than the energy resolution, stability of $E_F$ position is extremely important. The $E_F$ position was determined by measuring the gold Fermi edge and fitting it to the Gaussian-broadened Fermi-Dirac (FD) distribution function assuming a linear density of state. We evaluated stability of the $E_F$ position by repeated measurements of the Fermi edge. Figure S6 shows the difference of the deduced $E_F$ position at each channel of detector angle between two independent measurements with an interval of $\sim 1$ hour, and the results for three sets of measurements are shown. Since the $E_F$ position was independently determined for $\sim 350$ channels of detector angle, we deduced the average $\Delta E_F^{avg}$ and standard deviation $\Delta E_F^{\sigma}$ of those values as follows,

$$\Delta E_F^{avg} = \frac{1}{n} \sum_i \Delta E_F(i)$$

$$(\Delta E_F^{\sigma})^2 = \frac{1}{n} \sum_i (\Delta E_F(i) - \Delta E_F^{avg})^2,$$

where $\Delta E_F(i)$ is the difference of the $E_F$ position between two independent measurements with an interval at $i$th detector channel and $n$ is the number of detector channels $\sim 350$. Table SI shows the thus deduced values of $\Delta E_F^{avg}$ and $\Delta E_F^{\sigma}$ for the three sets of measurements. From these results, we have concluded that stability of the $E_F$ position during the

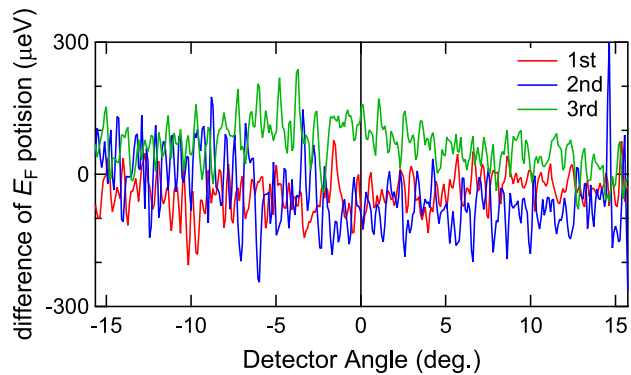

FIG. S6. Difference of $E_F$ position at each channel of detector angle between two independent measurements with an interval of $\sim 1$ hour. $E_F$ position at each channel was determined from the EDC of gold Fermi edge by fitting to the Gaussian-broadened FD function. Results for three sets of measurements are shown.

SC gap measurements is $\sim \pm\ 100\ \mu$eV.

TABLE. SI. $\Delta E_F^{avg}$ and $\Delta E_F^{\sigma}$ for three sets of measurements (see Supplemental text for detail).

|     | $\Delta E_F^{avg}$ | $\Delta E_F^{\sigma}$ |
|-----|--------------------|----------------------|
| 1st | -42 $\mu$eV        | 73 $\mu$eV           |
| 2nd | -47 $\mu$eV        | 104 $\mu$eV          |
| 3rd | 66 $\mu$eV         | 51 $\mu$eV           |